\documentclass{iopart}
\newcommand{\abs}[1]{\left|#1\right|}
\begin{document}
\title{Radiation-balanced simulations for binary inspiral}
\author{John T Whelan\dag\footnote[2]{Electronic Address:
    whelan@oates.utb.edu}, Christopher Beetle\S, {Walter Landry\S} and
  Richard H Price\S}
\address{\dag Department of Physics and Astronomy, University of Texas
  at Brownsville, Brownsville, TX 78520}
\address{\S Department of Physics, University of Utah,
         Salt Lake City, UT 84112}

\begin{abstract}
  The late stage of the inspiral of two black holes may have important
  non-Newtonian effects that are unrelated to radiation reaction. To
  understand these effects we approximate a slowly inspiralling binary
  by a stationary solution to Einstein's equations in which the holes
  orbit eternally. Radiation reaction is nullified by specifying a
  boundary condition at infinity containing equal amounts of ingoing
  and outgoing radiation. The computational problem is then converted
  from an evolution problem with initial data to a boundary value
  problem. In addition to providing an approximate inspiral waveform
  via extraction of the outgoing modes, our approximation can give
  alternative initial data for numerical relativity evolution.  We
  report results on simplified models and on progress in building 3D
  numerical solutions.
\end{abstract}

\submitto{\CQG}
\maketitle

\section{Introduction}

A pair of compact objects (black holes or neutron stars) in binary
orbit will, according to Newtonian gravity, remain in the same orbits
forever.  In general relativity, however, the system will emit
gravitational radiation, causing the bodies to spiral in towards one
another.  The gravitational radiation given off by this system is a
prime candidate for detection by upcoming gravitational wave
telescopes such as LIGO and VIRGO \cite{gwdet}.

Full three-plus-one numerical evolution of the Einstein equations is a
powerful tool for determining the history of such a system and the
form of the gravitational radiation emitted, but is limited in its
applicability by instabilities which prevent current codes from
evolving for more than an orbit or so. Thus it is not
only efficient but also necessary to limit full-numerical simulations
to the portion of the evolution for which there is not an applicable
approximation scheme \cite{laz}. The early stages of inspiral can be
handled by post-Newtonian methods \cite{pn}, while the final
post-merger ringdown can be treated with black hole perturbation
theory \cite{qn}.  The purpose of the quasi-stationary
approximation is to provide improved early waveforms and later-time
supercomputer initial conditions by modelling a late inspiral phase,
in which some nonperturbative gravitational effects are relevant, but
the radiation-induced inspiral is still occurring slowly.

\section{Quasi-Stationary Approximation}\label{sec:qsapprox}
  
The idea, initially proposed by Blackburn and Detweiler \cite{det},
is that if the inspiral is slow, the system is nearly periodic: after
one orbit, the objects have returned almost to their original
locations, and radiation which has moved out has been replaced with
new radiation of approximately the same shape.  If the objects' orbits
are circular rather than elliptical, the spacetime is nearly
stationary.  If the approximate orbital frequency is $\Omega$, moving
forward in time by $\delta t$ and rotating the resulting spatial slice
by $-\Omega\,\delta t$ will not change the picture very much.

This approximation can be used to simplify the numerical problem by
solving for an exactly stationary spacetime which serves, over some
period of time, as a reasonable approximation for the slowly evolving
spacetime.  In the process, the three-plus-one dimensional evolution
problem is reduced to a three-dimensional, nondynamical one, not only
reducing the size of the computational grid, but also hopefully
avoiding evolutionary instabilities.

Since gravitational radiation emitted from the orbiting system carries
away energy, some modification must be made to the physical problem in
order to allow this equilibrium solution.  One approach
\cite{confflat} is to require the spatial geometry of the
stationary spacetime to be conformally flat and solve for the
conformal factor using only a subset (the constraints) of the Einstein
equations.  This method enforces stationarity by in effect discarding
degrees of freedom considered to be radiative, which would carry
away energy.  Our approach, on the other hand, is to keep the
gravitational radiation, but nullify the radiation reaction by
balancing the outgoing radiation by an equal amount of incoming
radiation.  In so doing, we will solve the full Einstein equations in
the presence of the actual sources, and simply replace the physical
boundary condition of outgoing radiation at large distances with one
of balanced radiation.

\section{Radiation-Balanced Boundary Conditions}\label{sec:rbbc}

In the general context of a theory in which a pair of orbiting sources
for a wave equation give off radiation, it is instructive to consider
three types of solutions.  In a \textbf{Type~I} solution, the
radiation is outgoing, carrying away energy and causing the orbits to
decay.  In a \textbf{Type~II} solution, the radiation is again
outgoing, but some additional force, uncoupled to the radiating field,
keeps the sources in their circular orbits, resulting in a stationary
spacetime.  A \textbf{Type~III} solution is also stationary, with the
sources remaining in circular orbits of a constant frequency, but this
time there is no external force, and the equilibrium is maintained by
a balance of incoming and outgoing radiation.  Type~I is the physical
situation we ultimately wish to model.  Type~II should be a reasonable
approximation to Type~I if the Type~I solution in question is
inspiralling only slowly.  It should not be appropriate to general
relativity, in which all matter and energy couples to the
gravitational field, but is useful for comparison in theories which
allow for external forces.  Type~III is the stationary solution we
wish to find numerically, and, where appropriate, to relate to a
Type~II or I solution.

\subsection{Scalar Field Theory Results}\label{ssec:scalar}

To analyze in detail the implications of boundary conditions on
radiating, stationary solutions (Types II and III), prior work on this
project \cite{Whel00} has considered the theory of a nonlinear scalar
field $\psi(t,r,\theta,\phi)$ in Minkowski spacetime, the simplest
theory with nonlinearity, orbits, and radiation.  If the source is
required to rotate at a constant angular velocity $\Omega$, so that
the charge density $\rho(t,r,\theta,\phi)$ is a function only of $r$,
$\theta$, and $\varphi=\phi-\Omega t$, there exist stationary solutions
for which the field exhibits the same symmetry; in that case, the wave
equation can be written
\begin{equation}
  \label{eq:corot}
  \nabla^2-\Omega^2 c^{-2}
  \partial_\varphi^2 \psi
  = -\rho + \lambda \mathcal{F}(\psi) 
  ,
\end{equation}
where $\nabla^2$ is the spatial Laplacian and $\mathcal{F}(\psi)$
is a nonlinear coupling.  Convenient choices of charge distribution
include equal and opposite point charges at antipodal points in the
equatorial plane, or translationally invariant line charges
perpendicular to the equatorial plane.  (The latter choice renders the
problem equivalent to that of point charges in 2+1 dimensions.)

In the context of a numerical solution on a finite grid, the Type~II
(purely outgoing radiation) solution $\psi^R_{\mathrm{out}}$ is
defined by applying a Sommerfeld ($\psi_{,r}+c^{-1}\psi_{,t} =
\psi_{,r}-\Omega c^{-1}\psi_{,\varphi}=0$) boundary condition at the
radius $R$ of the outer boundary of the grid.  If the nonlinear term
becomes small at large distances, the form of any solution near a
large-$R$ boundary will be a solution to the vacuum, linear version of
(\ref{eq:corot}); each angular Fourier mode will be a linear
combination of two independent solutions, which can be identified as a
purely ingoing and a purely outgoing solution.  The limit
$\psi_{\mathrm{out}} := \lim_{R\rightarrow\infty}
\psi^R_{\mathrm{out}}$ is simply the solution consisting only of
outgoing modes, with the co\"{e}fficients of all the ingoing modes set
to zero.  [The entire discussion can be repeated for ingoing
radiation, with the resulting solution being
$\psi_{\mathrm{in}}(r,\theta,\varphi)
=\psi_{\mathrm{out}}(r,\theta,-\varphi)$.]

The most familiar Type~III radiation-balanced (RB) solutions are
standing-wave solutions in which the Dirichlet ($\psi=0$) or Neumann
($\psi_{,r}=0$) boundary condition is enforced at $r=R$.  If the
large-distance form of any of these solutions is resolved in angular
Fourier modes, it is found that the amplitudes of the co\"{e}fficients
of the ingoing and outgoing components are equal; however the relative
phase of the two independent vacuum solutions depends on the choice of
$R$, as does the radiation amplitude corresponding to a given source
strength.  Unlike outgoing- (or ingoing-) wave solutions, these
standing-wave solutions do not tend towards any limit as the boundary
radius is taken to infinity.

Waves satisfying Neumann or Dirichlet boundary conditions are not the
only solutions with equal magnitudes of incoming and outgoing
radiation, and in fact are not the type that are appropriate for our
problem. In the case of a linear theory ($\lambda=0$) the relevant
standing wave solution is simply a linear superposition of the ingoing
and outgoing solutions (LSIO) $\psi_{\mathrm{LSIO}}
=\frac{1}{2}(\psi_{\mathrm{in}}+\psi_{\mathrm{out}})$. For the
nonlinear problem we want an analog of this superposition. To this end
we note that in the linearized theory, the LSIO solution is the RB
solution with the minimum radiation in the wave zone. This means, in
effect, that the LSIO is the solution with 
just the radiation needed to keep the sources in equilibrium and no
``extra'' radiation.  From the point of view of using a Type~III
solution to approximate a Type~II one, we can take this ``minimum
energy radiation balance'' (MERB) solution, extract its outgoing
component, and identify it with the outgoing radiation.

In the nonlinear theory, the superposition of two solutions is no
longer a solution.  In \cite{Whel00}, we reformulated (\ref{eq:corot})
(in 2+1 dimensions, or equivalently with infinite line charges) as a
Green's function problem and using the average of the advanced and
retarded Green's functions to find a radiation-balanced solution.  We
found that even for highly nonlinear theories, this time-symmetric
Green's function method can be used to find a good approximation for
the outgoing solution.  The success of the approximation means that
superposition of the ingoing and outgoing solutions is approximately
valid even when nonlinear effects are strong. The explanation for 
this is crucial to the physical basis for our approximation: In the
innermost regions the fields are strong, but are very insensitive to
the boundary conditions; the fields here differ only slightly when
outgoing boundary conditions are changed to ingoing boundary conditions.
In the wave zone region, the fields depend very sensitively on the
boundary conditions, but here the fields are weak.  Thus superposition
approximately works because the solutions being superposed are almost
the same where nonlinearities are strong, and are almost linear where
the solutions are very different. The success of the numerical
verification of this principle lends confidence that in the case of
general relativity, if we can find a Type~III solution with little or
no ``superfluous'' radiation, we may be able to relate it to the
physical Type~I solution in an analogous way.

As an aside, it is worth stressing that while the equation
(\ref{eq:corot}) is elliptical inside and hyperbolic outside the
``speed of light cylinder'' $r\sin\theta=1/\Omega$, we experienced no
difficulties finding the numerical solution to the problem using
closed-surface boundary conditions usually associated with a purely
elliptical equation.  In particular there were no discernable
artefacts at the light cylinder, which required no special treatment in
the numerical solution.

\subsection{A more general prescription}\label{ssec:genrbbc}

Theories like GR, which have more involved nonlinearities than
(\ref{eq:corot}) cannot be reduced to the linear Green's function
problem described in Sec.~\ref{ssec:scalar}.  It is therefore
necessary to develop a more general method for both imposing the
condition of radiation balance and selecting a MERB solution.

Far from the sources, (\ref{eq:corot}) reduces to
\begin{equation}
  \label{eq:corotlim}
  r^{-1} \partial_r^2 (r\psi_\ell^m) + m^2 \Omega^2 \psi_\ell^m \rightarrow 0
\end{equation}
where we have resolved $\psi(r,\theta,\varphi) = \sum_{\ell
  m}\psi_\ell^m(r)Y_\ell^m(\theta,\varphi)$ in spherical harmonics.
The general solution in the wave zone will then be
\begin{equation}
  \label{eq:wavesoln}
  \psi_\ell^m \rightarrow r^{-1}(A_\ell^m e^{im\Omega r}
  +B_\ell^m e^{-im\Omega r})
  \ .
\end{equation}
Since $Y_\ell^m(\theta,\varphi)$ includes a factor of $e^{im\varphi}$
and hence $e^{-im\Omega t}$, the first term in (\ref{eq:wavesoln})
represents an outgoing solution and the second one an ingoing one.
The condition of radiation balance is that, for each $(\ell,m)$ mode,
$A_\ell^m$ and $B_\ell^m$ have the same amplitude.  In other words, we
can write $A_\ell^m/B_\ell^m=\exp(2i\delta_l^m)$, where $\delta_l^m$ is
an unspecified phase.  (The possible choices of $\delta_l^m$
parameterize the radiation-balanced family of solutions.)  Combining
this with the form (\ref{eq:wavesoln}) of the solution in the wave
zone, a particular radiation-balanced solution will obey
\begin{equation}
  \label{eq:radbalderiv}
  \frac{\partial_r\psi_\ell^m}{\psi_\ell^m}=im\Omega
  \frac{\exp[i(m\Omega r+\delta_\ell^m)] - \exp[-i(m\Omega r+\delta_\ell^m)]}
  {\exp[i(m\Omega r+\delta_\ell^m)] + \exp[-i(m\Omega r+\delta_\ell^m)]}
  \ .
\end{equation}

The family of boundary conditions (\ref{eq:radbalderiv}) allows us to
construct the family of radiation-balanced solution.  To find the MERB
solution, we just need to vary the $\delta_\ell^m$ to minimize the
total ``energy'' $2\sum_{\ell m}\abs{A_{\ell m}}^2$.

The actual numerical solution of (\ref{eq:corot}) is a boundary value
problem outside a mixed hyperbolic-elliptic region. The standing-wave
boundary conditions for the nonlinear problem, furthermore, are very
intricate since they require a three step process of multipole
decomposition, application of the boundary condition to each
multipole, and multipole summation. This multipole
decomposition/recomposition is easier to implement as one step of an
iterative relaxation process. Unfortunately {\em standard} relaxation
techniques will only work for a problem that is purely elliptic inside
the boundary. This is especially unfortunate since relaxation methods
are better suited to handling the numerically intensive work that will
be necessary for our gravitational three-dimensional problem. For that
reason we have developed a modified relaxation method that is not
limited to elliptic problems. This method has been applied to the
gravitational test problem of a single black hole.  The numerical
solution in fact has been found to converge correctly, but quite
slowly.  A major focus of the project now is improved numerical
algorithms for the boundary value problem.

\section{Stationary Rotating Solutions in General Relativity}

In the theory of a scalar field in Minkowski space, the stationarity
condition obeyed by Type II and III solutions was that the field
depended on the co\"{o}rdinates $t$ and $\phi$ only in the combination
$\varphi=\phi-\Omega t$, or equivalently, that $(\partial_t +
\Omega\partial_\phi)\psi=0$.  In general relativity, this condition
becomes a Killing symmetry $\mathcal{L}_K g_{ab}=0$, where the Killing
vector can be thought of as $K\sim\partial_t +\Omega\partial_\phi$.

An elegant way to impose the Killing symmetry is the Geroch
decomposition \cite{geroch}, in which the four-geometry is described
in terms of a set of scalar and vector fields on a three-dimensional
manifold of Killing vector trajectories, plus the three-metric of this
manifold.  However, since our Killing vector is timelike near the axis
of rotation and spacelike far from the axis (becoming null at the
light cylinder), the manifold of Killing vector orbits has a surface
of signature change, which makes the three-manifold of Killing vector
orbits a problematic starting point for numerical simulation of this
spacetime.

The approach currently being pursued is instead to perform the simulation in
harmonic co\"{o}rdinates
$\{x^i=t,x,y,z\}$, general analogues to Cartesian co\"{o}rdinates,
which are annihilated by the spacetime d'Alembertian when treated as
scalar fields.  In analogy to rotating co\"{o}rdinates in flat space,
we define $\xi = x\cos\Omega t + y\sin\Omega t$ and $\eta =
-x\sin\Omega t + y\cos\Omega t$, and requires that $(\partial/\partial
t)_{\xi,\eta,z}$ be a Killing vector.  This means that in
$\{t,\xi,\eta,z\}$ co\"{o}rdinates, the metric components only depend
on $\xi$, $\eta$, and $z$.  The induced functional dependence of the
metric components in the harmonic co\"{o}rdinate system is more
complicated, but ultimately the unknown functions for which we
numerically solve depend on only three co\"{o}rdinates.  The Einstein
equations, along with the harmonic gauge condition, can then be solved
on the hypersurface $t=0$, which completely determines the stationary
spacetime.

\section{Conclusions and Outlook}

This paper has described an ongoing research program to find
numerically a stationary spacetime which can approximate over some
stretch of time a slowly inspiralling compact object binary.  A
stationary solution to the radiative problem is to be achieved by
replacing the physical boundary condition of purely outgoing radiation
with a condition corresponding to an equal mix of ingoing and outgoing
radiation.

Numerical work so far has mostly focussed on the properties of
radiation-balanced solutions in scalar field theories (extracting
outgoing radiation as well as techniques for selecting a preferred
radiation-balanced solution in the first place), but progress is being
made towards implementing the method for the gravitational problem.
In the meantime, it is instructive to consider how such an equilibrium
radiating spacetime could be used.

First, since the spacetime will contain at large distances a
superposition of ingoing and outgoing gravitational radiation, we
should be able to separate out the outgoing-wave contribution and use
it as an approximation for the purely-outgoing radiation from the
physical system.

Second, a timelike hypersurface of the spacetime could be used as an
alternative to post-Newtonian or conformally flat initial data for a
full numerical evolution, in one of two ways.  First, while the RB
spacetime is an equilibrium solution of Einstein's equations, valid
for all time, it could still be used to provide initial data for a
full numerical evolution, since the outer boundary condition of such
an evolution would be one of outgoing rather than balanced radiation,
and thus once the ``missing'' ingoing radiation had failed to
propagate from the outer boundary to the location of the sources, the
orbits would begin to decay.  However, since that would mean consuming
precious supercomputer evolution waiting for the incoming waves to
stop, it would preferable to extend the method of extracting a metric
configuration containing only outgoing radiation to cover not just the
wave zone, but the entire time slice.

\ack


The authors wish to thank J.~Romano, W.~Krivan, S.~Detweiler,
P.~Brady, T.~Creighton, \'{E}.~Flanagan, S~.Hughes, K.~Thorne,
A.~Wiseman, C.~Torre, J.~Friedmann, S.~Morsink, A.~Held,
P.~H\'{a}j\'\i\v{c}ek, J.~Bi\v{c}\'{a}k, J.~Friedman, J.~Novak,
J.~Baker, B.~Br\"{u}gmann, M.~Campanelli, C.~Lousto, B~Whiting and
K.~Blackburn.  This work was University of Utah by the National
Science Foundation under grant PHY-9734871, and JTW was supported at
the University of Bern by the Swiss Nationalfonds, and by the Tomalla
Foundation, Z\"{u}rich, and at the University of Texas at Brownsville
by the National Science Foundation under grant PHY-9981795.

\end{document}